\documentclass{PoS}
\usepackage{url} 

\title{$CP$ violation in charm decays at Belle}

\ShortTitle{$CP$ violation in charm decays at Belle}

\author{\speaker{Byeong Rok Ko}\thanks{on behalf of the Belle Collaboration.}\\
        Korea University, SEOUL, Republic of Korea\\
        E-mail: \email{brko@hep.korea.ac.kr}}

\abstract{Using the full data sample collected with the Belle detector
  at the KEKB asymmetric-energy $e^+e^-$ collider, we present $CP$
  violation in charm decays. The $D^0-\bar{D}^0$ mixing parameter
  $y_{CP}$ and indirect $CP$ violation parameter $A_{\Gamma}$ in
  $D^0\rightarrow h^+h^-$ decays are reported, where $h$ denotes $K$
  and $\pi$. The preliminary results are
  $y_{CP}=(1.11\pm0.22\pm0.11)\%$ and
  $A_{\Gamma}=(-0.03\pm0.20\pm0.08)\%$. We also report searches for
  $CP$ violation in $D^0\rightarrow h^+h^-$ and $D^+\rightarrow K^0_S
  K^+$ decays. No evidence for $CP$ violation in $D^0\rightarrow
  h^+h^-$ is observed with $A^{KK}_{CP}=(-0.32\pm0.21\pm0.09)\%$ and
  $A^{\pi\pi}_{CP}=(+0.55\pm0.36\pm0.09)\%$. The $CP$ asymmetry
  difference between $D^0\rightarrow K^+K^-$ and
  $D^0\rightarrow\pi^+\pi^-$ decays is measured with $\Delta
  A^{hh}_{CP}=(-0.87\pm0.41\pm0.06)\%$. The $CP$ asymmetry in
  $D^+\rightarrow K^0_S K^+$ decay is measured to be
  $(-0.25\pm0.28\pm0.14)\%$. After subtracting $CP$ violation due to
  $K^0-\bar{K}^0$ mixing, the $CP$ asymmetry in
  $D^+\rightarrow\bar{K}^0 K^+$ decay is found to be
  $(+0.08\pm0.28\pm0.14)\%$.}

\FullConference{The European Physical Society Conference on High Energy Physics -EPS-HEP2013\\
		18-24 July 2013\\
		Stockholm, Sweden}

\begin{document}

\section{Introduction}
Violation of the combined Charge-conjugation and Parity symmetries
($CP$) in the standard model (SM) is produced by a non-vanishing phase
in the Cabibbo-Kobayashi-Maskawa flavor-mixing matrix~\cite{CKM} and
that in charm decays is expected to be very small in the
SM~\cite{SMCP, YAY}, thus it provides a unique probe to search for
beyond the SM.

\section{$y_{CP}$ and $A_{\Gamma}$ measurements with $D^0\rightarrow
  h^+h^-$ and $D^0\rightarrow K^-\pi^+$ decays}
The neutral charmed meson mixing and indirect $CP$ violation ($CPV$)
parameters, $y_{CP}$ and $A_\Gamma$ are defined as
\begin{eqnarray}
  \label{EQ:YCP}
  y_{CP}
  &=&\frac{\hat\Gamma(D^0\rightarrow
    h^+h^-)+\hat\Gamma(\bar{D}^0\rightarrow h^+h^-)}{2\Gamma}-1,\\
  \label{EQ:AGAMMA}
  A_{\Gamma}
  &=&\frac{\hat\Gamma(D^0\rightarrow h^+h^-)-\hat\Gamma(\bar{D}^0\rightarrow h^+h^-)}{2\Gamma},
\end{eqnarray}
where $\Gamma$ is the average decay width of the two mass eigenstates
of the neutral charmed mesons and $\hat\Gamma$ is the effective decay
width of $D^0\to h^+h^-$ that can be described with a single
exponential form~\cite{BERGMANN}. Under $CP$ conservation, $y_{CP}$ is
$y$ that is $\Delta\Gamma/2\Gamma$ and characterizes the charm
mixing where $\Delta\Gamma$ is the decay width difference between the
two mass eigenstates of the neutral charmed mesons. Therefore, any
large deviation between $y_{CP}$ and $y$ strongly indicates $CPV$ in
charm decays. 

The experimental observable for $y_{CP}$ is the lifetime difference
between $D^0\rightarrow h^+h^-$ and $D^0\rightarrow K^-\pi^+$ states,
where the former is $CP$-even and the latter is an equal mixture of
$CP$-even and $CP$-odd under $CP$ conservation. The $CPV$ parameter
$A_\Gamma$ can be measured from lifetime difference between the two
$CP$ conjugate decays. From Eq.~(\ref{EQ:YCP}) the lifetime of
$D^0\rightarrow h^+h^-$ can be expressed as $\tau(D^0\rightarrow
h^+h^-)=\tau/(1+y_{CP})$ and from (\ref{EQ:AGAMMA}) that of
$D^0\rightarrow h^+h^-$ and $\bar{D}^0\rightarrow h^+h^-$ can be
described with $\tau(D^0\rightarrow h^+h^-)=\tau(1-A_\Gamma)$ and
$\tau(\bar{D}^0\rightarrow h^+h^-)=\tau(1+A_\Gamma)$, respectively,
where $\tau$ is the lifetime of $D^0\rightarrow K^-\pi^+$. Therefore,
the lifetimes of $D^0\rightarrow h^+h^-$ and $\bar{D}^0\rightarrow
h^+h^-$ can be parameterized in terms of $y_{CP}$, $A_\Gamma$, and
$\tau$ as shown in Eq.~(\ref{EQ:TAUHH}).
\begin{eqnarray}
  \nonumber
  \tau(D^0\rightarrow h^+h^-)
  &=&\tau(1-A_\Gamma)/(1+y_{CP}),\\
  \label{EQ:TAUHH}    
  \tau(\bar{D}^0\rightarrow h^+h^-)
  &=&\tau(1+A_\Gamma)/(1+y_{CP}).
\end{eqnarray}
In order to extract $y_{CP}$, $A_\Gamma$, and $\tau$, we perform
simultaneous fit to the five proper decay time distributions from
$D^0\rightarrow K^+K^-$, $\bar{D}^0\rightarrow K^+K^-$,
$D^0\rightarrow K^-\pi^+$ + c.c., $D^0\rightarrow\pi^+\pi^-$, and
$\bar{D}^0\rightarrow\pi^+\pi^-$.
 
\begin{figure}[htb]
\begin{center}
\mbox{
  \includegraphics[width=0.9\textwidth,height=0.3\textwidth]{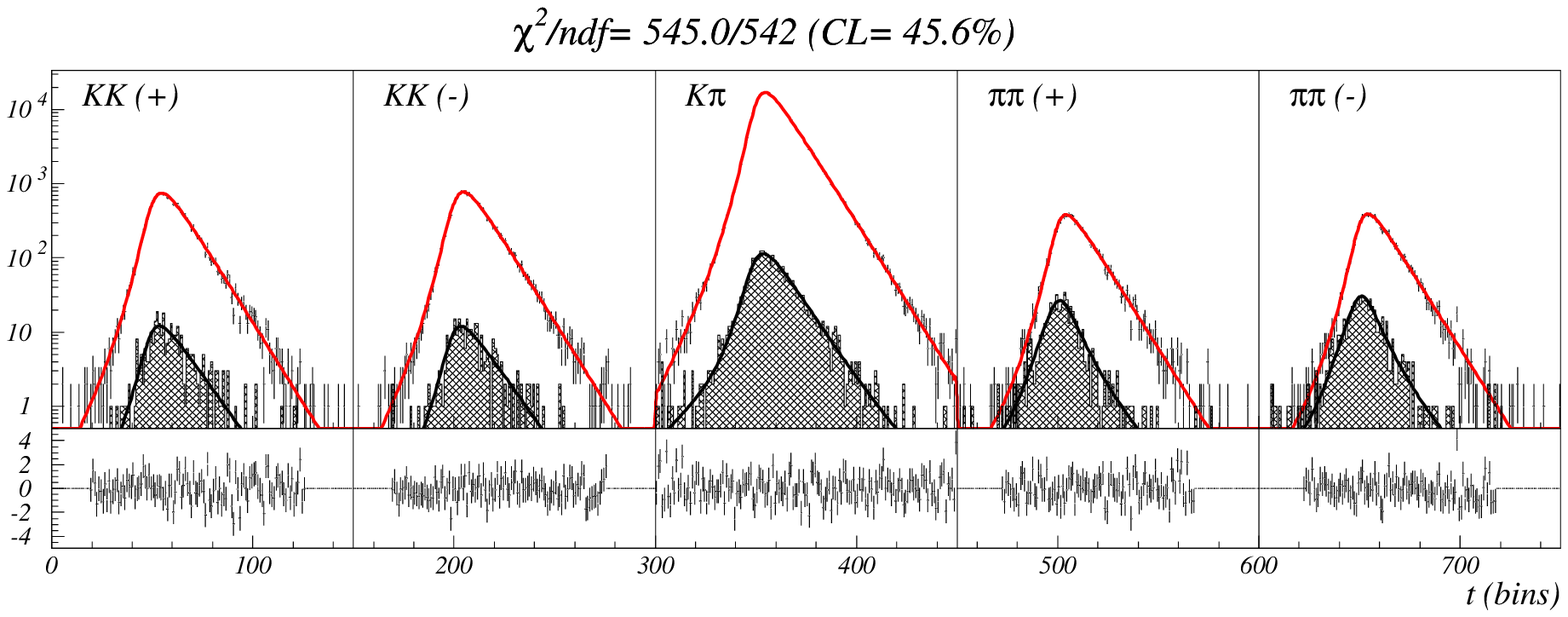}
}
\mbox{
  \includegraphics[width=0.9\textwidth,height=0.3\textwidth]{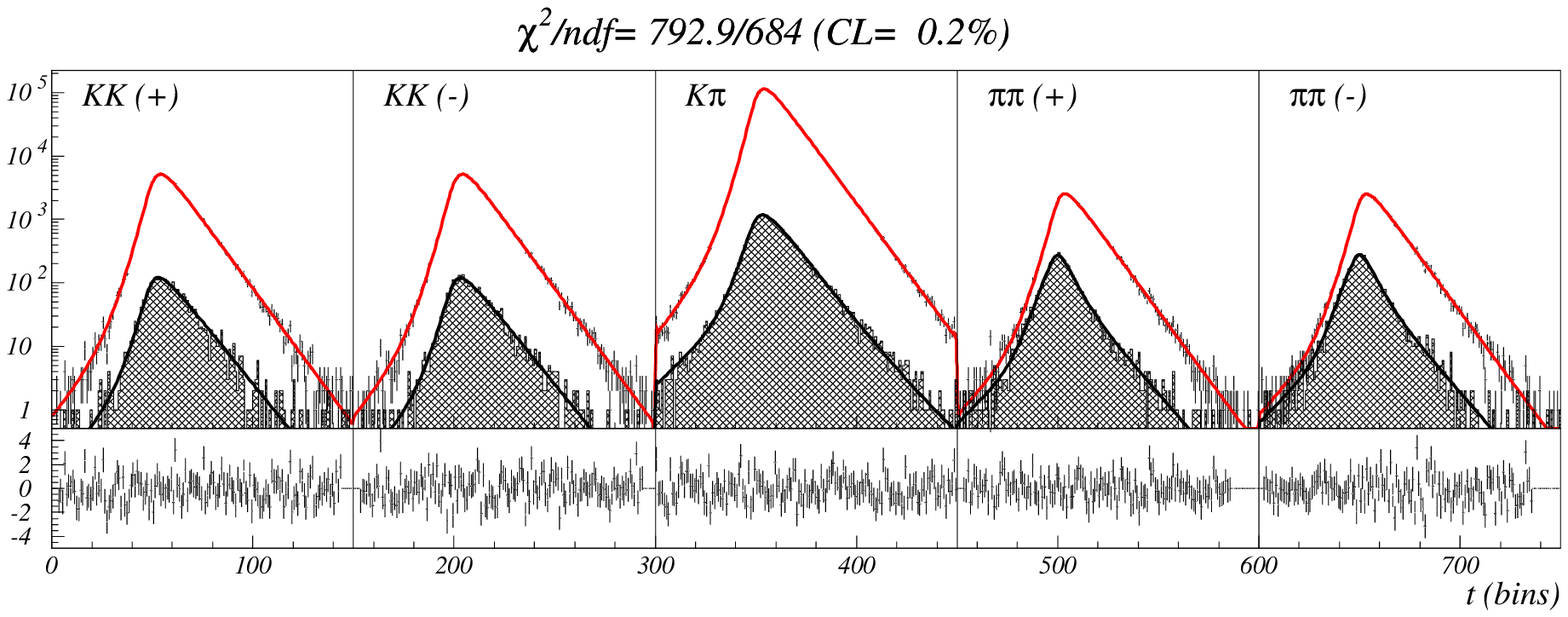}
}
\caption{Simultaneous fits to the proper decay time distributions that
  are integrated over the $\cos\theta^*$. Top (bottom) plots are
  obtained with 3-layer (4-layer) silicon vertex detector. The
  distributions of signal and sideband regions are shown as error bars
  and the hatched, respectively. The ``(+)'' and ``(-)'' denote the
  charge of the tagging soft pion.}
\label{FIG:YCP_FIT}
\end{center}
\end{figure}

\begin{figure}[htb]
\begin{center}
  \includegraphics[width=0.9\textwidth]{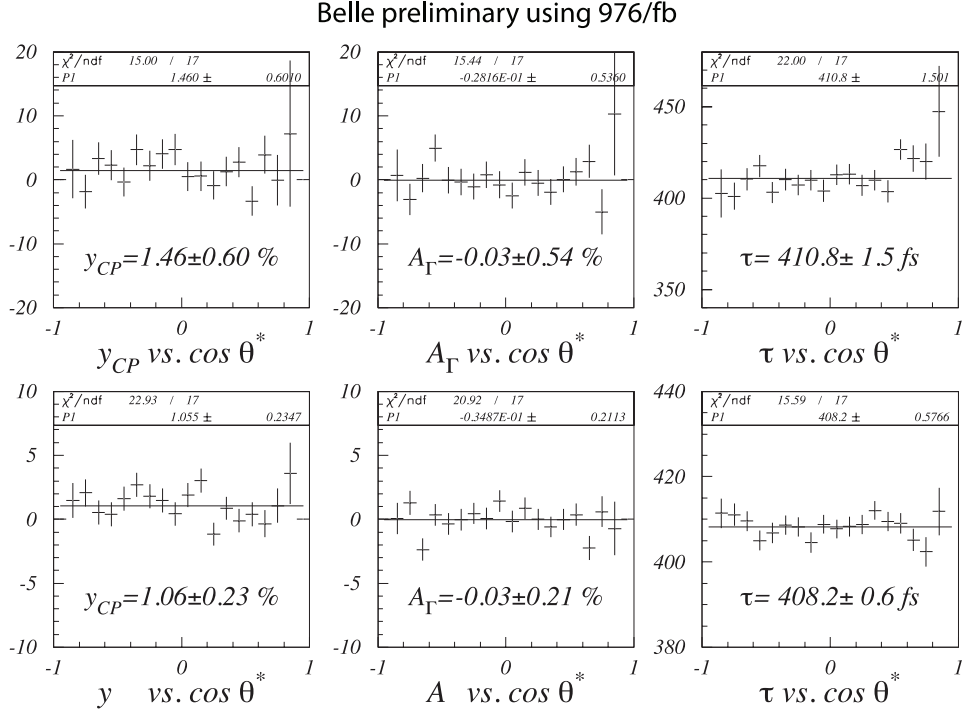}
\caption{$y_{CP}$, $A_\Gamma$, and $\tau$ as a function of the
  $\cos\theta^{*}$. Top (bottom) three plots are obtained with 3-layer
  (4-layer) silicon vertex detector.}
\label{FIG:YCP}
\end{center}
\end{figure}

Since the experimental data were taken with two different silicon
vertex detector configurations~\cite{SVD2}, we treat them separately
with the two different proper decay time resolution
functions. Figure~\ref{FIG:YCP_FIT} shows the simultaneous fits to the
five proper decay time distributions. To reduce systematic effects due
to the resolution function dependence on $\cos\theta^*$, where
$\theta^*$ is the polar angle of the $D^0$ momentum at the
center-of-mass system (c.m.s.), the simultaneous fits are actually
performed in bins of $\cos\theta^*$ to extract $y_{CP}$, $A_\Gamma$
and $\tau$. Figure~\ref{FIG:YCP} shows the results of the simultaneous
fits, $y_{CP}$, $A_\Gamma$, and $\tau$ as a function of the
$\cos\theta^{*}$. The averages of the fit results shown in
Fig.~\ref{FIG:YCP} are $y_{CP}=(1.11\pm0.22\pm0.11)\%$,
$A_\Gamma=(-0.03\pm0.20\pm0.08)\%$, and $\tau=(408.56\pm0.54)$ fs,
where the last is consistent with world average~\cite{PDG2012}. 

To conclude, we observe $y_{CP}$ with 4.5$\sigma$ significance and
find no indirect $CPV$ in $D^0\rightarrow h^+h^-$ decays.

\section{Direct $CPV$ measurements in $D^0\rightarrow h^+h^-$ and
  $D^+\rightarrow K^0_S K^+$ decays}
The direct $CP$ asymmetry of $D\to f$ decays is defined as
\begin{equation}
A^{D\rightarrow f}_{CP}=\frac{\Gamma(D\to
  f)-\Gamma(\bar{D}\to\bar{f})}{\Gamma(D\to
  f)+\Gamma(\bar{D}\to\bar{f})},
\end{equation}
where $\Gamma$ is the partial decay width. Experimental determination
of $A^{D\to f}_{CP}$ can be done with the asymmetry in the signal
yield
\begin{equation}
A^{D\rightarrow f}_{\rm rec}=\frac{N^{D\to f}_{\rm
    rec}-N^{\bar{D}\to\bar{f}}_{\rm rec}}{N^{D\to f}_{\rm
    rec}+N^{\bar{D}\to\bar{f}}_{\rm rec}}=A^{D\to f}_{CP} + A_{\rm others},
\end{equation}
where $N_{\rm rec}$ is the number of reconstructed decays and $A_{\rm
  others}$ are asymmetries other than $A^{D\to f}_{CP}$, production
and particle detection asymmetries. The methods developed in
Refs.~\cite{KSHPRL} and \cite{D0HHPLB} are used to correct for charged
kaon and soft pion detection asymmetries, respectively. To correct for
asymmetry caused by neutral kaons, we rely on the method in
Ref.~\cite{K0MAT}. Once we correct for asymmetries due to particle
detection, then we extract $A^{D\to f}_{CP}$ using the antisymmetry of
the production asymmetry which is the forward-backward asymmetry at
Belle.

\begin{figure}[htb]
\begin{center}
\mbox{
  \includegraphics[width=0.65\textwidth]{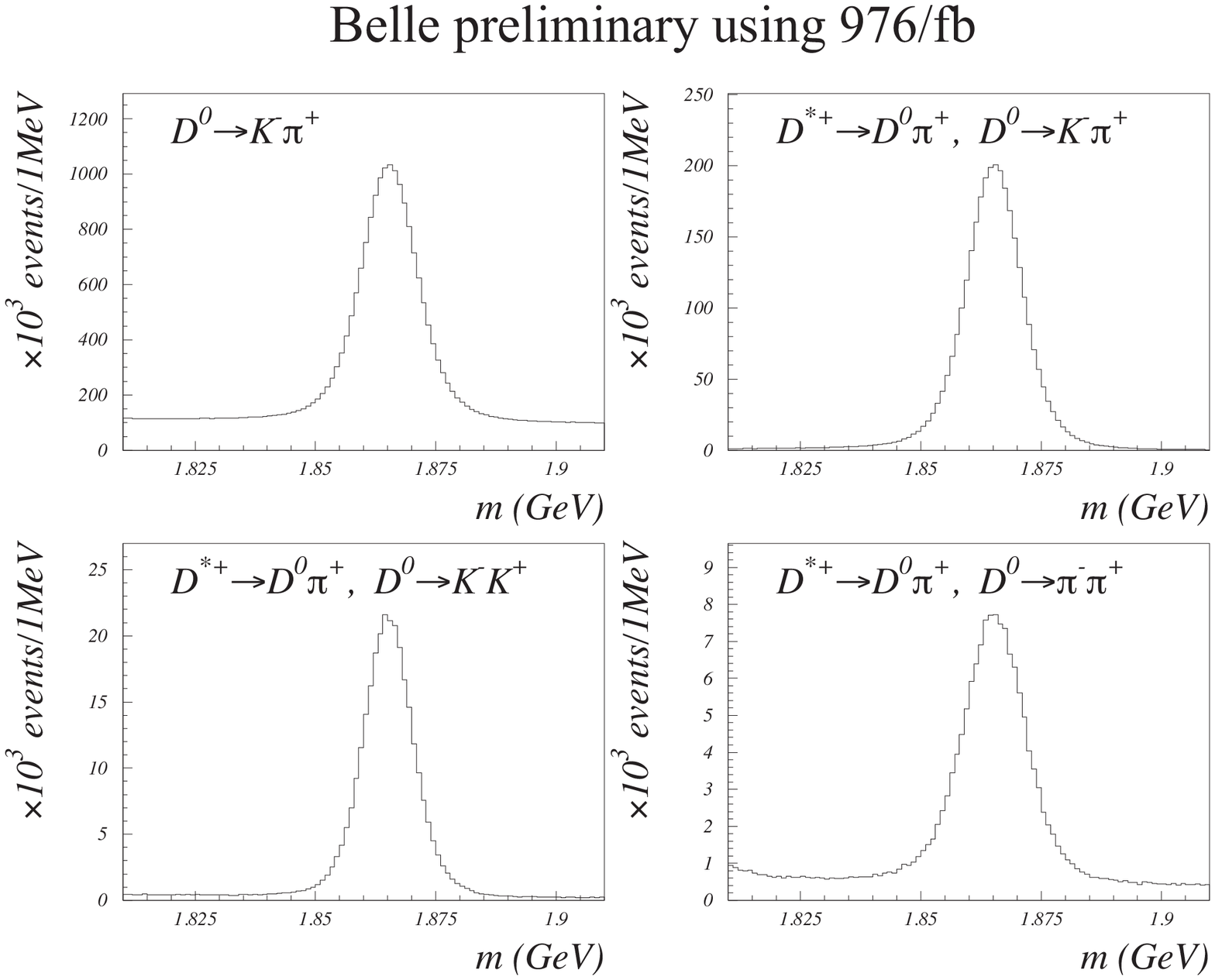}
}
\mbox{
  \includegraphics[width=0.65\textwidth]{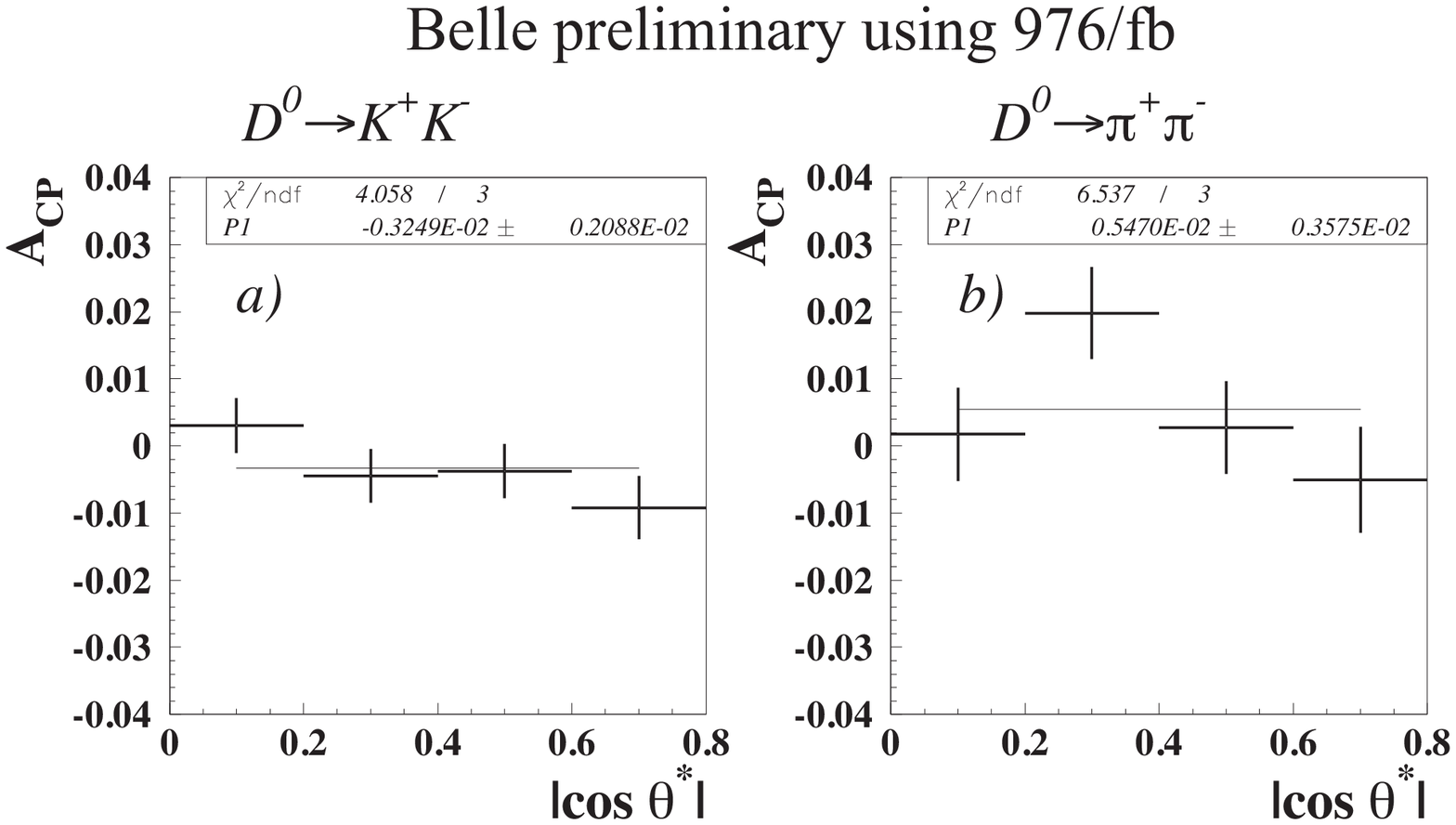}
}
\caption{Top four plots show reconstructed signal distributions
  described in the text and bottom two plots show preliminary results
  of $A_{CP}$ as a function of the polar angle of $D^{*+}$ momentum at
  the c.m.s.}
\label{FIG:ACPD0hh}
\end{center}
\end{figure}

The $D^0\rightarrow h^+h^-$ final states are singly Cabibbo-suppressed
(SCS) decays in which both direct and indirect $CPV$ are
expected in the SM~\cite{SMCP, YAY}, while the $CP$ asymmetry
difference between the two decays, $\Delta
A^{hh}_{CP}=A^{KK}_{CP}-A^{\pi\pi}_{CP}$ reveals approximately direct
$CPV$ with the universality of indirect $CPV$ in
charm decays~\cite{YAY}. Figure~\ref{FIG:ACPD0hh} shows reconstructed
signal distributions showing 14.7M $D^0\rightarrow K^-\pi^+$, 3.1M
$D^{*+}$ tagged $D^0\rightarrow K^-\pi^+$, 282k $D^{*+}$ tagged
$D^0\rightarrow K^+K^-$, and 123k $D^{*+}$ tagged
$D^0\rightarrow\pi^+\pi^-$, respectively, and the measured $A_{CP}$ in
bins of $|\cos\theta^{*}_{D^{*+}}|$. From the bottom plots in
Fig.~\ref{FIG:ACPD0hh}, we obtain
$A^{KK}_{CP}=(-0.32\pm0.21\pm0.09)\%$ and
$A^{\pi\pi}_{CP}=(+0.55\pm0.36\pm0.09)\%$ where the former shows the
best sensitivity to date. From the two measurements, we obtain $\Delta
A^{hh}_{CP}=(-0.87\pm0.41\pm0.06)\%$.

\begin{figure}[htb]
\begin{center}
\mbox{
  \includegraphics[width=0.50\textwidth]{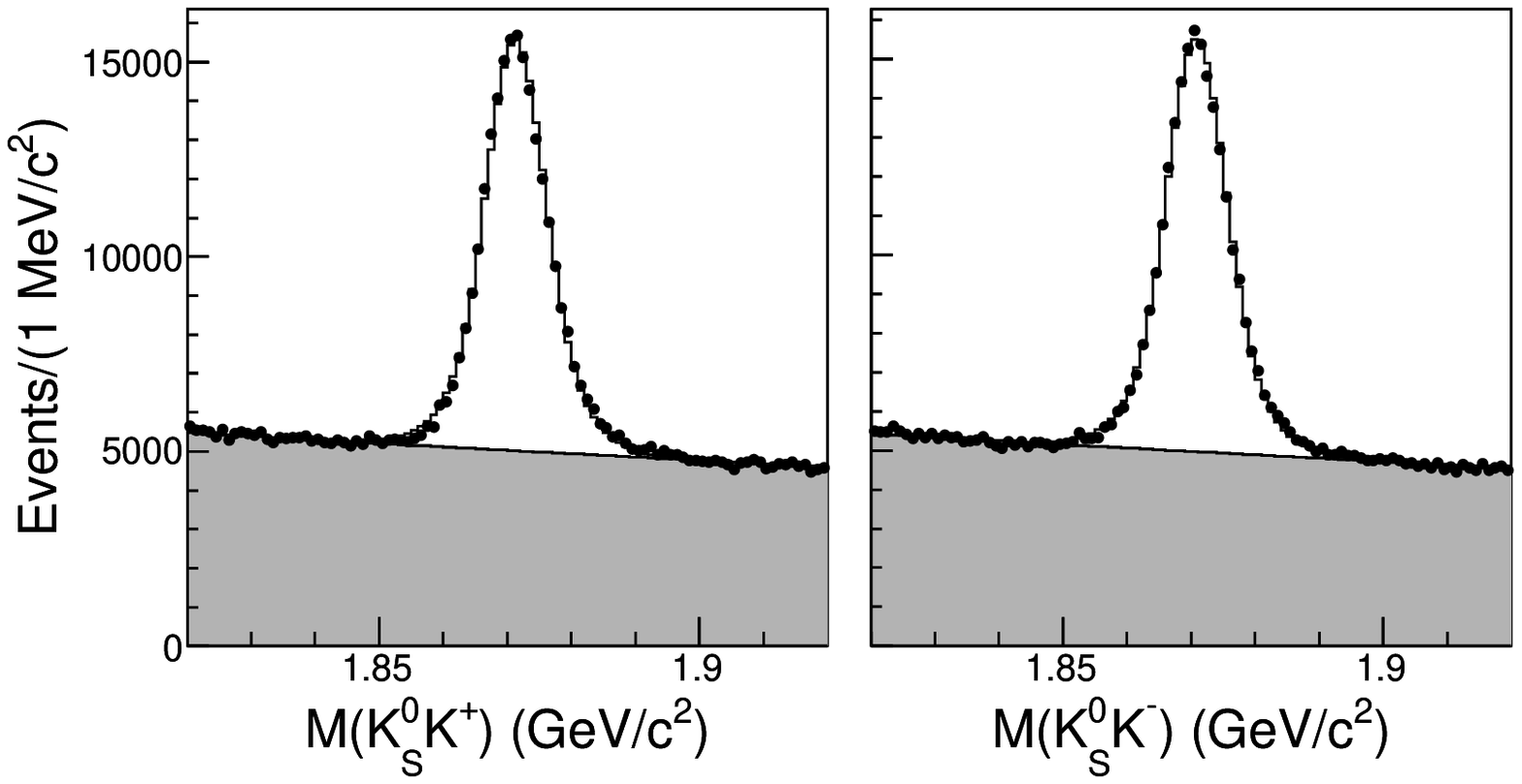}
}
\mbox{
  \includegraphics[width=0.45\textwidth]{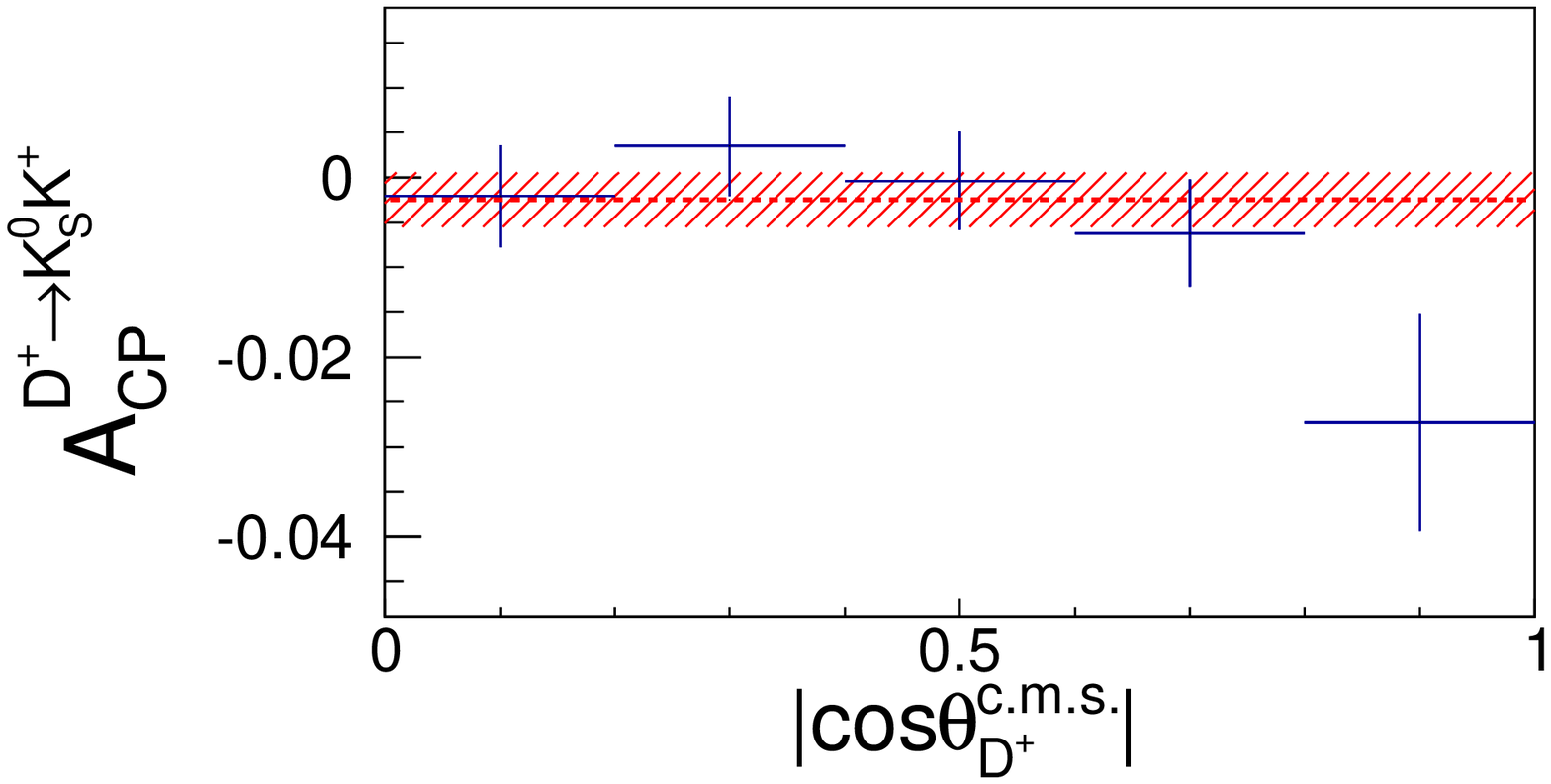}
}
\caption{Left two plots show $M(K^0_SK^+)$ and $M(K^0_SK^-)$
  distributions, respectively, and right plot shows $A_{CP}$ in the
  decay as a function of the polar angle of $D^{+}$ momentum at the
  c.m.s.}
\label{FIG:ACPKSK}
\end{center}
\end{figure}

The $D^+$ decaying to the final state $K^0_S K^+$ proceeds from
$D^+\rightarrow\bar{K}^0K^+$ decay which is SCS, where direct $CPV$ is
predicted to occur~\cite{SMCP, YAY}. With a $K^0_S$ in the final
state, $D^+\rightarrow K^0_S K^+$ decay is also expected to generate
$CPV$ due to $K^0-\bar{K}^0$ mixing, referred to as
$A^{\bar{K}^0}_{CP}$. The decay $D^+\rightarrow\bar{K}^0 K^+$ shares
the same decay diagrams with $D^0\rightarrow K^+K^-$ by exchanging the
spectator quarks, $d\leftrightarrow u$. Therefore, neglecting the
helicity and color suppressed contributions in
$D^+\rightarrow\bar{K}^0 K^+$ and $D^0\rightarrow K^+K^-$ decays, the
direct $CPV$ in the two decays is expected to be effectively the
same. Thus, as a complementary test of the $\Delta A^{hh}_{CP}$
measurement\footnote{Now the tension is rather
  released~\cite{HFAG_NOW}, but was strong~\cite{HFAG_OLD}.}, the
precise measurement of $A_{CP}$ in $D^+\rightarrow\bar{K}^0 K^+$ helps
to pin down the origin of $\Delta
A^{hh}_{CP}$~\cite{BBACPKSK}. Figure~\ref{FIG:ACPKSK} shows invariant
masses of $D^{\pm}\rightarrow K^0_S K^{\pm}$ together with the fits
that result in $\sim$277k reconstructed decays and the measured
$A_{CP}$ in bins of $|\cos\theta^{\rm c.m.s.}_{D^+}|$. From the right
plot in Fig.~\ref{FIG:ACPKSK}, we obtain $A^{D^+\rightarrow K^0_S
  K^+}_{CP}=(-0.25\pm0.28\pm0.14)\%$. After subtracting experiment
dependent $A^{\bar{K}^0}_{CP}$~\cite{GROSSMAN_NIR}, the $CPV$ in charm
decay, $A^{D^+\rightarrow\bar{K}^0 K^+}_{CP}$, is measured to be
$(+0.08\pm0.28\pm0.14)\%$~\cite{KSKJHEP}.

\section{Summary}
In summary, using the full data sample collected with the Belle detector
  at the KEKB asymmetric-energy $e^+e^-$ collider, we report the charm
  mixing parameter $y_{CP}$ and indirect $CPV$ parameter $A_{\Gamma}$
  using $D^0\rightarrow h^+h^-$ and $D^0\rightarrow K^-\pi^+$
  decays. The preliminary results are:
\begin{eqnarray}
  \nonumber
  y_{CP}&=&(1.11\pm0.22\pm0.11)\%,\\
  \nonumber
  A_{\Gamma}&=&(-0.03\pm0.20\pm0.08)\%.
\end{eqnarray}
We also report searches for $CP$ violation in $D^0\rightarrow h^+h^-$
and $D^+\rightarrow K^0_S K^+$ decays. The preliminary results of
$A_{CP}$ in $D^0\rightarrow h^+h^-$ decays and the difference between
the two $A_{CP}$ results are:
\begin{eqnarray}
  \nonumber
  A^{KK}_{CP}&=&(-0.32\pm0.21\pm0.09)\%,\\
  \nonumber
  A^{\pi\pi}_{CP}&=&(+0.55\pm0.36\pm0.09)\%,\\
  \nonumber
  \Delta A^{hh}_{CP}&=&(-0.87\pm0.41\pm0.06)\%,
\end{eqnarray}
and the results of $A_{CP}$ in $D^+\rightarrow K^0_S K^+$ decays are:
\begin{eqnarray}
  \nonumber
  A^{D^+\rightarrow K^0_S K^+}_{CP}&=&(-0.25\pm0.28\pm0.14)\%,\\
  \nonumber
  A^{D^+\rightarrow\bar{K}^0 K^+}_{CP}&=&(+0.08\pm0.28\pm0.14)\%.
\end{eqnarray}


\begin{thebibliography}{99}
\bibitem{CKM}
M. Kobayashi and T. Maskawa, Prog. Theor. Phys., {\bf 49}, 652 (1973).

\bibitem{SMCP}
F. Buccella {\it et al.}, Phys. Rev. D {\bf 51}, 3478 (1995).

\bibitem{YAY}
Y. Grossman, A. L. Kagan, and Y. Nir, Phys. Rev. D {\bf 75}, 036008 (2007).

\bibitem{BERGMANN}
S. Bergmann {\it et al.}, Phys. Lett. B {\bf 486}, 418 (2000).

\bibitem{SVD2} 
Z. Natkaniec {\it et al.} (Belle SVD2 Group), Nucl. Instr. and Meth. A
{\bf 560}, 1 (2006); Y. Ushiroda (Belle SVD2 Group), Nucl. Instr. and
Meth. A {\bf 511}, 6 (2003).

\bibitem{PDG2012}
J. Beringer {\it et al.} (Particle Data Group), Phys. Rev. D {\bf 86},
010001 (2012).

\bibitem{KSHPRL} 
B. R. Ko {\it et al.} (Belle collaboration), Phys. Rev. Lett. {\bf 104}, 181602 (2010).

\bibitem{D0HHPLB} 
M. Stari\v{c} {\it et al.} (Belle collaboration), Phys. Lett. B {\bf 670}, 190 (2008).

\bibitem{K0MAT}
B. R. Ko, E. Won, B. Golob, and P. Pakhlov, Phys. Rev. D {\bf 84} 111501 (2011).

\bibitem{HFAG_NOW} 
\url{http://www.slac.stanford.edu/xorg/hfag/charm/March13/DCPV/direct_indirect_cpv.html}.

\bibitem{HFAG_OLD} 
\url{http://www.slac.stanford.edu/xorg/hfag/charm/ICHEP12/DCPV/direct_indirect_cpv.html}.

\bibitem{BBACPKSK}
B. Bhattacharya, M. Gronau, and J. L. Rosner, Phys. Rev. D {\bf 85}, 054104 (2012).

\bibitem{GROSSMAN_NIR} 
Y. Grossman and Y. Nir, JHEP. 04 ({\bf 2012}) 002.

\bibitem{KSKJHEP} 
B. R. Ko {\it et al.} (Belle Collaboration), JHEP. 02 ({\bf 2013}) 098.

\end{thebibliography}
\end{document}